\documentclass[preprint,showpacs,pre,12pt]{revtex4-1}
\usepackage{amsfonts}
\usepackage{amssymb}
\usepackage{amsmath}
\usepackage{graphicx}
\usepackage{pgfplots}

\begin{document}
\title{Random matrices associated with general barrier billiards}
\author{Eugene Bogomolny}
\affiliation{ Universit\'e Paris-Saclay, CNRS, LPTMS, 91405 Orsay, France}
\date{\today}

\begin{abstract}

The paper is devoted to the derivation   of  random unitary  matrices whose  spectral statistics is the same as statistics of quantum eigenvalues of certain deterministic two-dimensional barrier billiards. These random matrices are extracted from the exact billiard quantisation condition  by applying   a random phase approximation for high-excited states.  An important ingredient  of the method is the calculation of $S$-matrix for the scattering in the slab with a half-plane inside by the Wiener-Hopf method. It appears that these random matrices have the form similar to the one obtained by the author in [arXiv:2107.03364] for a particular case of symmetric barrier billiards but with different choices of parameters.  The local correlation functions of the resulting random matrices are  well approximated by the semi-Poisson distribution which is a characteristic feature of various models with intermediate statistics.  Consequently, local spectral statistics of the considered barrier billiards is (i) universal for almost all values of parameters and (ii)  well described  by the semi-Poisson statistics. 

\end{abstract}

\maketitle

\section{Introduction}\label{introduction}

Polygonal billiards constitute a special class of classical dynamical systems. Though they have zero Lyapunov exponents the behaviour of their trajectories is intricate and complicated due to, in general,  unavoidable  discontinuities of the ray dynamics (see, e.g., \cite{gutkin}). An important subset of polygonal billiards is constituted by the so-called pseudo-integrable billiards (see, e.g., \cite{richens_berry}) characterised by the requirement that all their angles $\theta_j$  are rational multiples of $\pi$
\begin{equation}
\theta_j=\frac{m_j}{n_j}\pi
\label{angles}
\end{equation}
with co-prime integers $m_j$ and $n_j$. A characteristic property of pseudo-integrable billiards is the fact that their trajectories cover surfaces of finite genus connected with angles  by the formula \cite{katok}
\begin{equation}
g=1+\frac{N_n}{2}\sum_{j} \frac{m_j-1}{n_{j}} 
\label{genus}
\end{equation}
where $N_n$ is the least common multiply of all denominators $n_j$. This is a clear-cut difference of pseudo-integrable billiards (with at least one $m_j>1$) from  both limiting cases of classical dynamical models: integrable models where trajectories belong to tori (i.e., surfaces with $g=1$) and chaotic models where a typical trajectory covers the whole surface of constant energy.  

It is plain that peculiarities of pseudo-integrable billiards should have quantum manifestations but no general statements about statistical properties of pseudo-integrable models have been proposed so far. It is in a strong contrast with quantum integrable and fully chaotic  models  where the well-known and well-accepted conjectures were established long time ago \cite{berry_tabor} and \cite{BGS}. Numerical results \cite{cheon}-\cite{gorin_wiersig} suggest that, at least, for certain pseudo-integrable billiards spectral statistics differs from both the Poisson statistics of  integrable models \cite{berry_tabor} as well as from the usual random matrix statistics conjectured for  chaotic motels \cite{BGS}. Surprisingly, the observed statistics (called intermediate statistics) is similar to the spectral statistics of the Anderson model at the point of metal-insulator transition \cite{altshuler, shklovskii} whose main features are (i) linear level repulsion as for usual random matrix ensembles and (ii) an exponential fall-off of nearest-neighbour distributions as for the Poisson distribution. 

\begin{figure}
\begin{minipage}{.49\linewidth}
\begin{center}
\includegraphics[width=.97\linewidth]{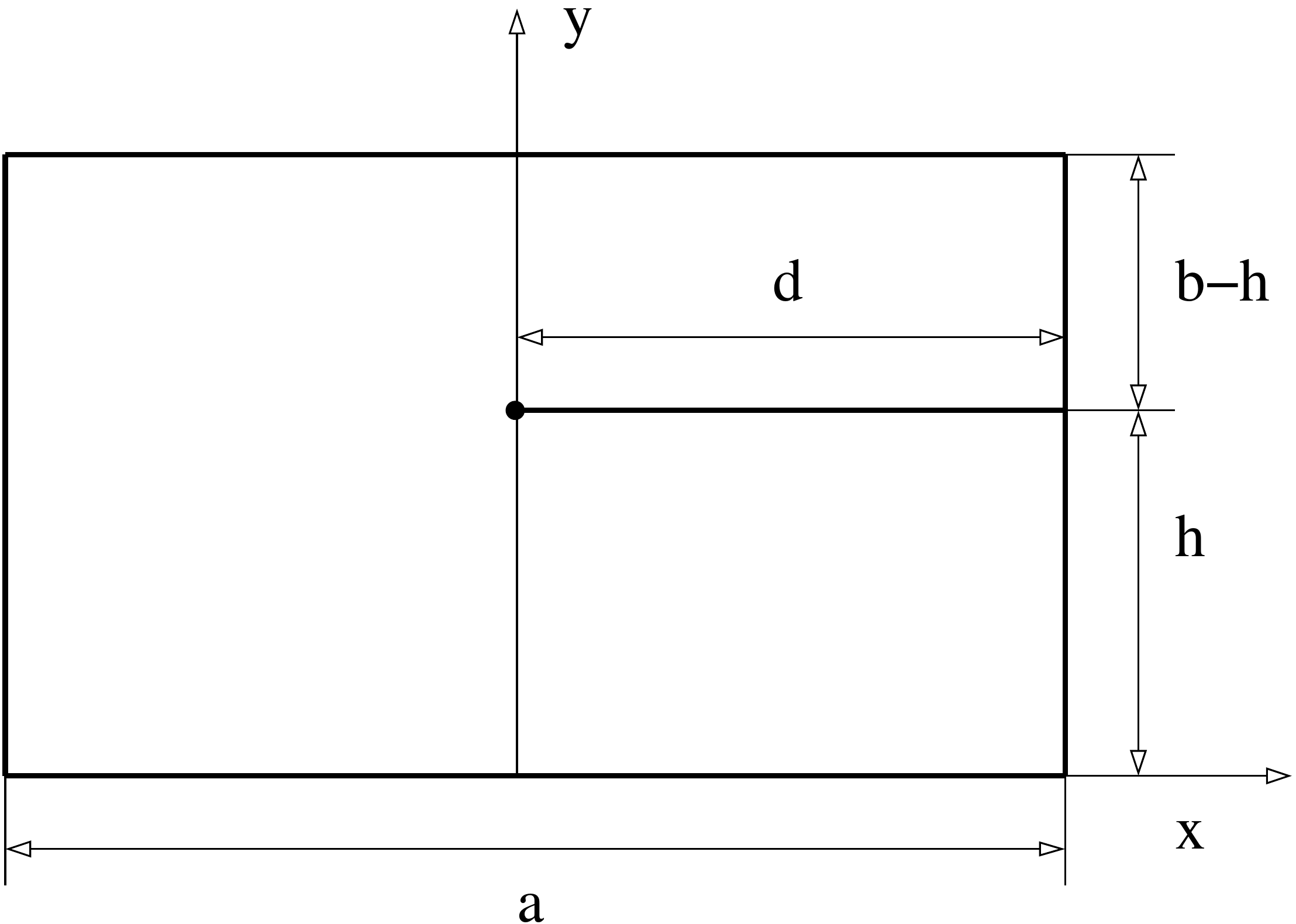}\\
(a)
\end{center}
\end{minipage}
\begin{minipage}{.49\linewidth}
\begin{center}
\includegraphics[width=.97\linewidth]{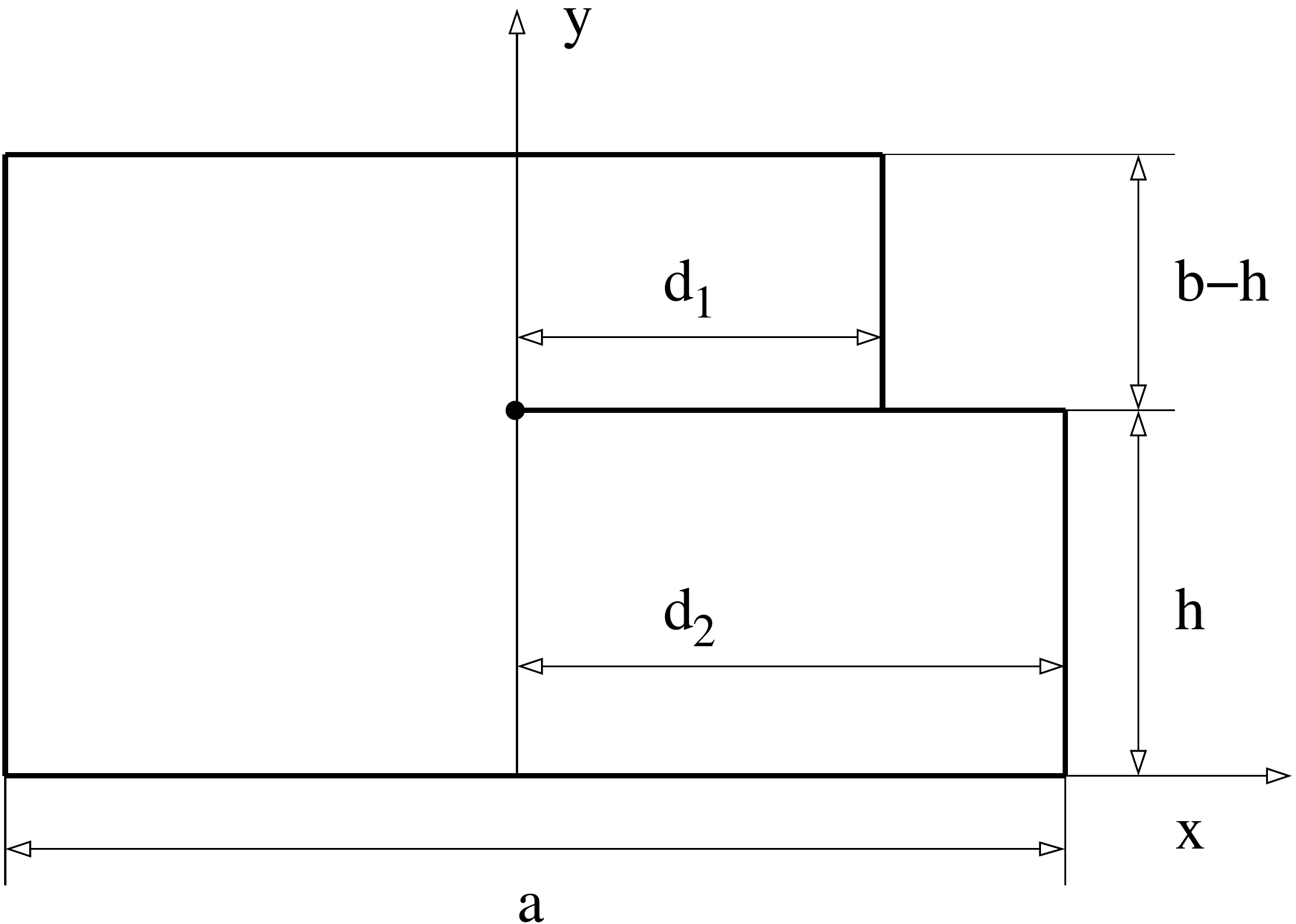}\\
(b)
\end{center}
\end{minipage}
\caption{Barrier billiards considered in the paper. For clarity the barrier tip is indicated by a small circle.}
\label{barrier}
\end{figure}

The simplest pseudo-integrable model is a rectangular billiard with a  barrier inside (see figure~\ref{barrier}(a)). It has 6  internal angles equal $\pi/2$ and one angle $2\pi$. According to \eqref{genus} it corresponds to a   genus 2 surface. 
Quantum problem for such billiards consists in finding eigenvalues $k$  and eigenfunctions $\Psi(x,y;k)$ of the Helmholtz equation
\begin{equation}
\left (\Delta+k^2 \right )\Psi(x,y;k)=0
\label{helmholtz} 
\end{equation}
provided that functions $\Psi(x,y;k)$ obey the Dirichlet boundary conditions 
\begin{equation}
\Psi(x,y;k)|_{\mathrm{boundaries}}=0.
\label{dirichlet}
\end{equation}
Generalisation for another type of boundary conditions is straightforward. 

Numerical calculations \cite{wiersig, gorin_wiersig} were done exclusively for a  symmetric barrier billiard  with $h=b/2$.  A new method of extracting random matrices from the exact quantisation of a symmetric barrier billiard has been proposed in \cite{bogomolny}. The main conclusion  of that paper  is that spectral statistics of symmetric barrier billiard is the same as the one for a $N\times N$ random unitary matrix
\begin{equation}
B_{\mu,\nu}=e^{i\Phi_{\mu} } \frac{ L_{\mu} L_{\nu}}{x_{\mu}+x_{\nu} },\qquad \mu,\nu=1,\ldots N
\label{random_matrix}
\end{equation}
where $\Phi_{\mu}$ are independent real random variables  distributed uniformly between $0$ and $2\pi$. $L_{\mu}$ are real quantities determined by the expression
\begin{equation}
L_{\mu}^2=2x_{\mu} \prod_{\nu \neq \mu}\frac{x_{\mu}+x_{\nu}}{x_{\mu}-x_{\nu} }
\label{L_m}
\end{equation} 
and coordinates $x_{\mu} $ for a symmetric barrier billiard (with $h=b/2$) are  
\begin{equation}
x_{\mu}=(-1)^{\mu+1}\sqrt{ k^2-\frac{\pi^2 \mu^2}{b^2}}\, .
\end{equation}
The condition that all $x_{\mu}$ are real defines matrix dimension $N$ 
\begin{equation}
N=\left[\frac{ kb}{\pi}\right ]
\end{equation}
where $[x]$ is the largest integer less or equal $x$.
 
The purpose of the paper is to generalise  the method of \cite{bogomolny} for non-symmetric barrier billiards shown  in figure~\ref{barrier}. It appears that the random matrix extracted from the exact quantisation condition for these billiards also have the form given by \eqref{random_matrix} and \eqref{L_m} but with a different specification of coordinates $x_{\mu}$.  

The plan of the paper is the following.  Section~\ref{sos} is devoted to the discussion of the surface-of-section method in application to barrier billiards  proposed in \cite{bogomolny}. The method consists in opening the billiard and finding exact scattering solutions for the resulting slab with a half-plane inside. Writing exact eigenfunctions of a barrier billiard as a linear combination of scattering waves and imposing the correct boundary conditions on previously removed parts of billiard boundary gives  the quantisation condition in the form $\det(1+B)=0$ where matrix $B$ differs form the $S$-matrix for the scattering in the slab by certain phase-like  factors. It is known that spectral statistics  of matrix $B$ is up to a rescaling coincides with statistics of high-excited barrier billiard eigenenergies \cite{semiclassics, DS}.   The main part of the paper consists in the  calculation of the exact scattering $S$-matrix by the Wiener-Hopf method. This is done in Section~\ref{non_resonant_S}. As it is typical for the Wiener-Hopf method the resulting expressions include infinite products and are quite cumbersome. In Section~\ref{unitary_random_matrix} two  important simplifications appeared in the semiclassical limit are discussed. First, by  ignoring  exponentially decreasing evanescent modes one gets finite dimensional unitary $S$ and $B$ matrices. Second,  the phase factors by which the $B$-matrix differs from the $S$-matrix are considered as independent random variables uniformly distributed on the unit circles.  Noticing that proper $S$-matrix  phases  by conjugation lead only to a shift  of random phases of the $B$-matrix   one proves that the resulting unitary random matrix has the form \eqref{random_matrix} and \eqref{L_m} but  a different  definition of coordinates $x_{\mu}$.  In has been argued in \cite{bogomolny} that local correlation functions of such matrices should be well described by the semi-Poisson distribution appeared in different models with intermediate statistics \cite{plasma_model}. This statement is illustrated by numerical calculations for certain typical barrier billiard parameters  in the end of Section~\ref{unitary_random_matrix}.  Section~\ref{summary} is a brief recapitulation  of the main steps permitting to extract random matrices from the exact solution of barrier billiard problems. Appendix~\ref{appendix_a} presents the factorisation of the kernel appeared in the Wiener-Hopf method and the explicit verification of the unitarity of the obtained scattering $S$-matrix.  
 

\section{Exact surface-of-section quantisation}\label{sos}

\begin{figure}
\begin{center}
\includegraphics[width=.5\linewidth]{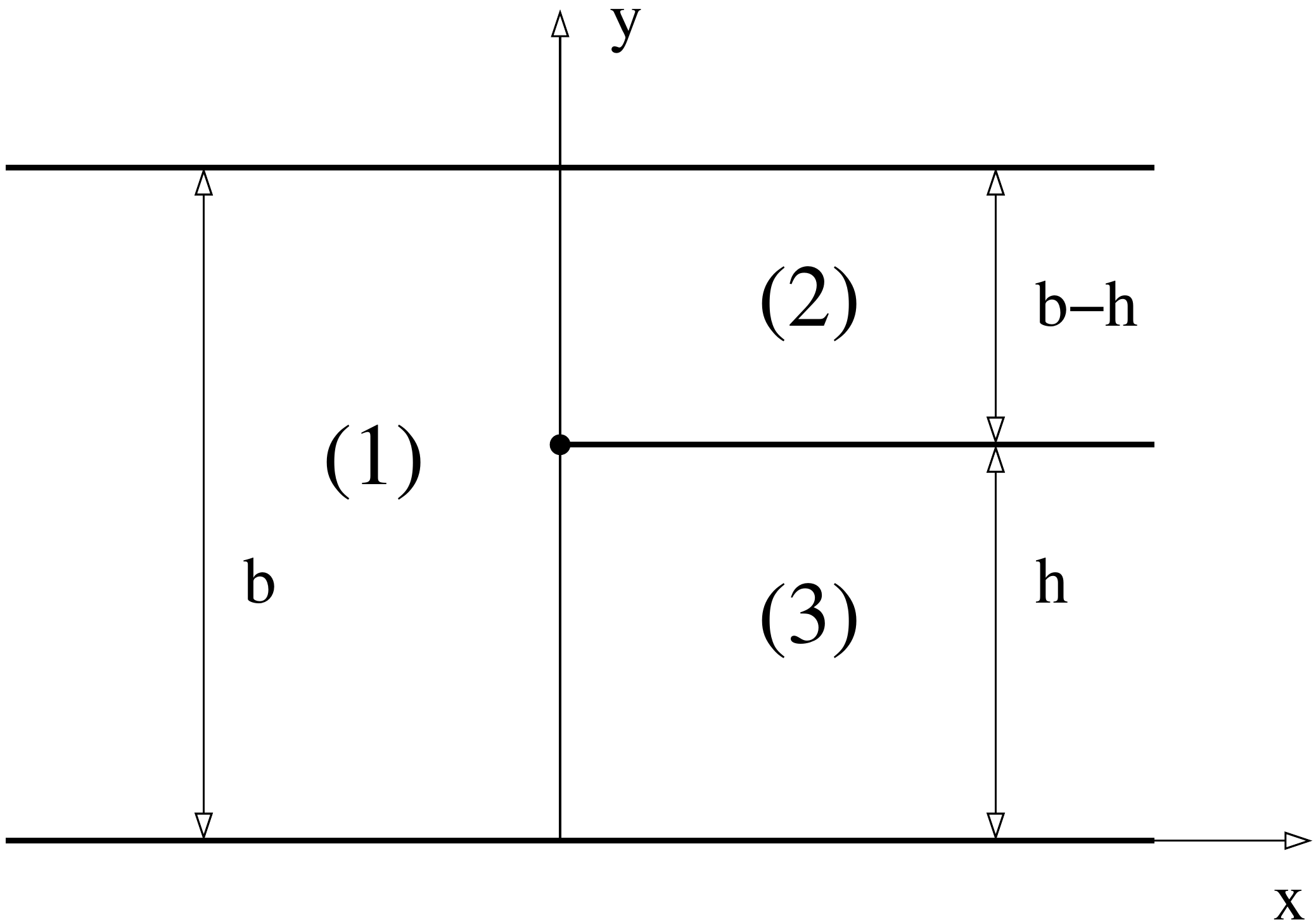}
\end{center}
\caption{An infinite  slab with a half-plane inside. Numbers indicate 3 possible channels.}
\label{general_barrier}
\end{figure}

The first step of the method proposed in \cite{bogomolny} consists in the removal  boundaries perpendicular to the barrier  and considering instead of a closed system the  problem of a scattering inside an infinite slab of height $b$ with a half-plane inside as  indicated in figure~\ref{general_barrier}. The Dirichlet boundary conditions \eqref{dirichlet} are imposed in all boundaries. 

Inside the slab there exit 3 different regions (channels) (see figure~\ref{general_barrier}). The following functions constitute elementary solutions in these channels (normalised to unit current)
\begin{equation}
\phi_n^{(\beta)\pm}(x,y)= \frac{e^{\pm ip_n^{(\beta)} x}}{\sqrt{ h_{\beta}  p_n^{(\beta)}}}\sin\Big (\frac{\pi n}{h_{\beta}}y\Big),\qquad p_n^{(\beta)}=\sqrt{k^2-\frac{\pi^2 n^2}{h_{\beta}^2}}, \qquad \beta=1,2,3 \, .
\label{elementary_solutions}
\end{equation}
Here $h_{\beta}$ is the width of channel $\beta$
\begin{equation}
h_1=b,\quad h_2=b-h,\quad h_3=h.
\label{h_beta}
\end{equation}
By definition, these functions are zero outside the corresponding channels. 
Subscript $(+)$ (resp., $(-)$)  indicates plane waves propagating from left to right (resp., from  right to left).  

We consider 3 different solutions corresponded to 3 possible plane waves entering from the infinity into the indicated regions of the slab.  They are denoted  by $\Psi_n^{(\beta)}(x,y)$ where superscript $(\beta)$ with $\beta=1,2,3$  indicates the entering channel and subscript $n$ is the transverse quantum number of the incident plane wave.
Their formal expansions into elementary solutions    \eqref{elementary_solutions}  are 
\begin{eqnarray}
\Psi_n^{(1)}(x,y)&=&\phi_{n}^{(1)+}(x,y)+\sum_{m=1}^{\infty} \left [ S_{n,m}^{1\to 1} \phi_{m}^{(1)-}(x,y)+
S_{n,m}^{1\to 2} \phi_{m}^{(2)+}(x,y)+S_{n,m}^{1\to 3} \phi_{m}^{(3)+}(x,y)\right ],
\label{the_first}\\
\Psi_n^{(2)}(x,y)&=&\phi_{n}^{(2)-}(x,y)+\sum_{m=1}^{\infty} \left [ S_{n,m}^{2\to 1} \phi_{m}^{(1)-}(x,y)+
S_{n,m}^{2\to 2} \phi_{m}^{(2)+}(x,y)+S_{n,m}^{2\to 3} \phi_{m}^{(3)+}(x,y)\right ],
\label{the_second}\\
\Psi_n^{(3)}(x,y)&=&\phi_{n}^{(3)-}(x,y)+\sum_{m=1}^{\infty} \left [ S_{n,m}^{3\to 1} \phi_{m}^{(1)-}(x,y)+
S_{n,m}^{3\to 2} \phi_{m}^{(2)+}(x,y)+S_{n,m}^{3\to 3} \phi_{m}^{(3)+}(x,y)\right ].
\label{the_third}
\end{eqnarray}
The first terms are the incident plane waves with momentum $p_n^{(\beta)}$ and the remaining sums represent the reflected and transmitted waves in all channels.

Coefficients $S_{n,m}^{\alpha\to\beta}$ with $\alpha,\beta=1,2,3$ form the $S$-matrix for the scattering inside the slab indicated in figure~\eqref{general_barrier}. In the next Section it will be  calculated analytically by the Wiener-Hopf method. 

When these coefficients are known,  functions \eqref{the_first}-\eqref{the_third} are solutions of the Helmholtz equation \eqref{helmholtz}  obeying the correct boundary conditions on all horizontal boundaries. Let us construct  an eigenfunction of barrier billiards indicated in figure~\ref{barrier} as a linear combination of all scattering waves
\begin{equation}
\Psi(x,y;k)=\sum_{n=1}^{\infty}\sum_{\alpha=1}^{3} a_n^{(\alpha)} \Psi_n^{(\alpha)}(x,y)\, .
\end{equation}
Coefficients $a_n^{(\alpha)}$ have to be determined from the requirement that this wave obeys the correct boundary conditions on vertical boundaries. For the general billiard as in figure~\ref{barrier}(b) these conditions are 
\begin{eqnarray}
\Psi(d_1,y;k)&=&0,\qquad h<y<b ,\label{d1}\\
\Psi(d_2,y;k)&=&0,\qquad 0<y<h, \label{d2}\\
\Psi(d_2-a,y;k)&=& 0,  \qquad 0<y<b.\label{a_d2}
\end{eqnarray}
In each channel function $\Psi_n(x,y)$ is a linear combination of elementary solutions \eqref{elementary_solutions} which form a complete and orthogonal set of functions of $y$. Therefore,   the Dirichlet boundary conditions \eqref{d1}-\eqref{a_d2} imply that the coefficients of these linear combinations have to be zero. It means that $a_n^{(\alpha)} $ are determined from the equations
\begin{equation}
a_m^{(\beta)}+e^{i\phi_m^{(\beta)}} \sum_{n=1}^{\infty}\sum_{\alpha=1}^3 a_n^{(\alpha)}S_{n,m}^{\alpha\to \beta}=0,\qquad \beta=1,2,3 \, .
\end{equation} 
Here quantities  $\phi_m^{(\alpha)}$ are 
\begin{equation}
\phi_m^{(1)}=2p_m^{(1)}(a-d_2),\qquad \phi_m^{(2)}=2p_m^{(2)}d_1,\qquad \phi_m^{(3)}=2p_m^{(3)}d_2\, .
\label{exact_phases}
\end{equation}
These relations can compactly be rewritten as follows (with the standard convention that the summation is performed over repeated indices)
\begin{equation}
a_m^{(\beta)}+a_n^{(\alpha)}B_{n,m}^{\alpha\to \beta}=0
\end{equation} 
where
\begin{equation}
B_{n,m}^{\alpha\to \beta}=e^{i\phi_m^{(\beta)}} S_{n,m}^{\alpha\to \beta}\, .
\label{B_matrix}
\end{equation}
Matrix $B$ has two groups of indices, $\alpha,\beta=1,2,3$ indicate the initial and final channels and indices $n,m $ are positive  integers denoted transverse quantum numbers of waves propagating in these channels.
 
The condition of compatibility of these equation (i.e., the quantisation condition on momentum $k$) is
\begin{equation}
\det(\delta_{m,n}\delta_{\alpha,\beta}+B_{n,m}^{\alpha\to \beta})=0\, .
\label{determinant}
\end{equation}


\section{Calculation of  scattering $S$-matrix}\label{non_resonant_S}

To use the formulas of the preceding Section it is necessary to know the $S$-matrix  which describes the scattering inside the slab with a half-plane inside (cf., figure~\ref{general_barrier}). A convenient (and probably the simplest) way to calculate it is to use the Wiener-Hopf method. It is an old,  powerful, and well known method of solving certain diffraction-like problems (see, e.g., \cite{noble} and references therein) by reduction them to a special equation (called the scalar Wigher-Hopf equation) of the following form 
\begin{equation}
K(\alpha)X_{+}(\alpha)+X_{-}(\alpha)=f(\alpha)\, .
\label{WH_equation}
\end{equation}
Here $X_{+}(\alpha)$ and $X_{-}(\alpha)$ are two unknown functions of complex variable $\alpha$. Functions $K(\alpha)$ and $f(\alpha)$ are supposed to be  known. 

The essence of the Wiener-Hopf method is the assumption that function $X_{+}(\alpha)$ is free from singularities in the upper half-plane  $\mathrm{Im}\,\alpha>-\epsilon_1$ 
and function $X_{-}(\alpha)$ has no singularities in the lower half-plane $\mathrm{Im}\,\alpha<\epsilon_2$
with $\epsilon_{1,2}>0$. These requirements are usually achieved by representing functions $X_{\pm}(\alpha)$ as one-sided Fourier transforms 
\begin{equation}
X_{+}(\alpha)=\int_0^{\infty} \psi(x)e^{i\alpha x}dx,\qquad 
X_{-}(\alpha)=\int_{-\infty}^0 \psi(x)e^{i\alpha x}dx
\end{equation}
where function $\psi(x)$ exponentially decays at  $|x|\to\infty$
\begin{equation}
\psi(x)\underset{x\to +\infty}{\longrightarrow} e^{-\epsilon_1 x},\qquad \psi(x)\underset{x\to -\infty}{\longrightarrow} e^{\epsilon_2 x  }. 
\end{equation}
The solution  the Wiener-Hopf equation \eqref{WH_equation}  consists in the following steps (see \cite{noble} for details and proofs).
\begin{itemize}
\item Factorise the kernel $K(\alpha)$ into a product of two functions $K(\alpha)=K_{+}(\alpha)K_{-}(\alpha)$ where functions $K_{\pm}(\alpha)$ are free of singularities and zeros in, respectively, upper and lower half-planes.
\item Divide the both parts of \eqref{WH_equation} by $K_{-}(\alpha)$
\begin{equation}
K_{+}(\alpha)X_{+}(\alpha)+\frac{X_{-}(\alpha)}{K_{-}(\alpha)}=\frac{f(\alpha)}{K_{-}(\alpha)}.
\end{equation}
\item Represent the right-hand side of this equation as a sum of functions $F_{\pm}(\alpha)$ free of singularities in the corresponding half-planes 
\begin{equation}
\frac{f(\alpha)}{K_{-}(\alpha)}=F_{+}(\alpha)+F_{-}(\alpha).
\label{f_K}
\end{equation}
\item After such transformation  the Wiener-Hopf equation becomes
\begin{equation}
\frac{X_{-}(\alpha)}{K_{-}(\alpha)}-F_{-}(\alpha)=F_{+}(\alpha)-K_{+}(\alpha)X_{+}(\alpha).
\end{equation}
\end{itemize}
By construction the left-hand side of this equation is free of singularities in the lower half-plane  $\mathrm{Im}\,\alpha<\epsilon_2$ and the right-hand side has no singularities in the upper half-plane $\mathrm{Im}\,\alpha >-\epsilon_1$. As these planes have a common part,  the both parts have to be free of singularities in the whole complex plane of $\alpha$. Therefore they have to be a polynomial in $\alpha$. Usually from the boundary conditions it follows that this polynomial is zero. It such case the solutions of the Wiener-Hopf equation are 
\begin{equation}
X_{-}(\alpha)=K_{-}(\alpha)F_{-}(\alpha),\qquad X_{+}(\alpha)=\frac{F_{+}(\alpha)}{K_{+}(\alpha)}\, . 
\label{WH_solutions}
\end{equation}
The reduction of a given problem to the Wiener-Hopf equation (when it is possible) can be done by different methods. Below we follow the one discussed in \cite{noble}. 

A solution of a scattering problem for the Helmholtz equation  consists of two parts, an incident wave and a reflected wave
\begin{equation}
\Psi(x,y)=\Psi_{\mathrm{inc}}(x,y)+\psi_{\mathrm{ref}}(x,y).
\end{equation}
Assume that momentum $k$ in \eqref{helmholtz} has a small positive imaginary part: $k\to k+i\delta k$.  Then it is known that the reflected wave tends to zero when $|x|\to\infty$ (this is the radiation condition). From expansions similar to \eqref{the_first}-\eqref{the_third} it follows that $\psi_{\mathrm{ref}}(x,y)\longrightarrow e^{-r\delta k|x|}$ with $r\geq 1$. Therefore one can take $\epsilon_1=\epsilon_2=\delta k$ and the following quantities
\begin{equation}
\Phi_{+}(\alpha,y)=\int_0^{\infty} \psi_{\mathrm{ref}}(x,y)e^{i\alpha x}dx,\qquad 
\Phi_{-}(\alpha,y)=\int_{-\infty}^0 \psi_{\mathrm{ref}}(x,y)e^{i\alpha x}dx
\end{equation} 
are free of singularities in, respectively, the upper half-plane $ \mathrm{Im}\,\alpha >-\delta k$ and the lower half-plane $\mathrm{Im}\,\alpha <\delta k $. 

As $\Psi_{\mathrm{inc}}(x,y)$ and $\psi_{\mathrm{ref}}(x,y)$ obey the Helmholtz equation \eqref{helmholtz} quantities  $\Phi_{\pm}(\alpha,y)$ have to obey the equation
\begin{equation}
\Big (\frac{\partial^2}{\partial y^2}+q^2(\alpha)\Big )\Phi_{\pm}(\alpha,y)=0, \qquad q^2(\alpha)=k^2-\alpha^2\, . 
\end{equation} 
Because there are no obstacles in the horizontal directions, 
the necessary solutions which are zero at horizontal boundaries of the slab are 
\begin{equation}
\Phi_{+}(\alpha,y)+\Phi_{-} (\alpha,y)=\left \{ \begin{array}{lc} 
A(\alpha)\sin \big (q(\alpha)y\big ), &0<y<h\\
B(\alpha)\sin \big (q(\alpha)(b-y)\big ), &h<y<b\end{array}\right .  .  
\label{A_B}
\end{equation}
When functions $A(\alpha)$ and $B(\alpha)$ are calculated, the reflection field is given by the inverse Fourier transformation
\begin{equation}
\psi(x,y)=\frac{1}{2\pi} \int_{-\infty}^{\infty} e^{-i\alpha x}d\alpha \left \{ \begin{array}{l c} 
A(\alpha) \sin\big (q(\alpha) y\big ), & 0<y<h \\ B(\alpha)\sin \big (q(\alpha)(b-y)\big ), & h<y<b \end{array}\right . . 
\label{inverse_fourier}
\end{equation}
For positive $x$ the integration contour can be shifted in the lower half-plane of $\alpha$ and for negative $x$ one can shift the contour into the upper half-plane. 

To find uniquely functions  $\Phi_{\pm}(\alpha,y)$ it is necessary to know their values  at the line $y=h$  which  follow from  general  properties of wave functions.
\begin{itemize}
\item At the half-line $y=h$, $0<x<\infty$ the total field has to be zero, $\Psi(x,h)=0$. It means that 
\begin{equation}
\int_{0}^{\infty}\Psi_{\mathrm{inc}}(x,h) e^{i\alpha x} dx+\Phi_{+}(\alpha, h)=0. 
\label{zero_h}
\end{equation}
\item The total field $\Psi(x,y)$ and its $y$-derivative $\Psi^{\prime}(x,y)$ have to be continuous at $y=h$ and negative $x$. The incident waves for all 3 solutions is continuous at $y=h$ but their $y$-derivatives have a jump for the second and third solutions.  It leads to the following relation ($ ^{\prime}$ indicates the derivative over $y$)
\begin{equation}
\int_{-\infty}^0 \left (\Psi_{\mathrm{inc}}^{\prime} (x,h+0)-\Psi_{\mathrm{inc}}^{\prime}(x,h-0)\right )e^{i\alpha x}dx+\left (\Phi_{-}^{\prime} (x,h+0)-\Phi_{-}^{\prime}(x,h-0)\right )=0.
\label{y_prime_h}
\end{equation} 
\end{itemize}
It appears that the 3 scattering solutions \eqref{the_first}-\eqref{the_third}  lead to  Eq.~\eqref{WH_equation} with the same kernel $K(\alpha)$ but with different right-hand sides. For completeness the calculations are sketched below. 
\subsection{First solution}
For this solution the incident field  $\Psi_{\mathrm{inc}}(x,y)=\phi_{n}^{(1)+}(x,y)$ (cf. \eqref{the_first}). From \eqref{zero_h} it follows that 
\begin{equation}
\Phi_{+}(\alpha,h)=-\frac{i \sin (\pi nh/b )}{\big (\alpha+p_n^{(1)}\big )\sqrt{bp_n^{(1)}}}\, .
\label{phi_plus}
\end{equation}
From \eqref{A_B} one gets
\begin{equation}
\Phi_{-}(\alpha,h)-\frac{i \sin (\pi nh/b )}{\big (\alpha+p_n^{(1)}\big )\sqrt{bp_n^{(1)}}}=A(\alpha)\sin\big (q(\alpha)h\big )=
B(\alpha)\sin\big (q(\alpha)(b-h)\big )\, .
\label{phi_minus}
\end{equation}
Calculation of the $y$-derivative of the total field at the both sides of the barrier $y=h$, $x>0$  gives the following equations 
\begin{eqnarray}
\Phi_{+}^{\prime}(\alpha, h+0)+\Phi_{-}^{\prime}(\alpha, h)&=&-q(\alpha)B(\alpha) \cos\big (q(\alpha)(b-h)\big ),\nonumber\\
\Phi_{+}^{\prime}(\alpha, h-0)+\Phi_{-}^{\prime}(\alpha, h)&=&q(\alpha)A(\alpha) \cos\big (q(\alpha)h\big ).
\end{eqnarray}
Denoting 
$
X_{+}(\alpha)=b\big (\Phi_{+}^{\prime}(\alpha, h+0)-\Phi_{+}^{\prime}(\alpha, h-0)\big )
$
one finds
\begin{equation}
X_{+}(\alpha) =-b q(\alpha) \Big (A(\alpha) \cos\big (q(\alpha)h\big )+B(\alpha) \cos\big (q(\alpha)(b-h)\big )\Big )\, .
\end{equation}
Using  \eqref{phi_minus} one can express  $A(\alpha)$ and $B(\alpha)$ through $\Phi_{-}(\alpha,h)\equiv X_{-}(\alpha)$.  It   gives the final Wiener-Hopf equation \eqref{WH_equation} with 
\begin{equation}
K(\alpha)=\frac{\sin\big (q(\alpha)h\big )\sin\big (q(\alpha)(b-h)\big )}{b q(\alpha)\sin\big (q(\alpha)b\big )}
\label{K_alpha}
\end{equation}
and
\begin{equation}
f_1(\alpha)=\frac{i \sin (\pi nh/b )}{\big (\alpha+p_n^{(1)}\big )\sqrt{bp_n^{(1)}}}.
\label{WH_first_solution}
\end{equation}
The necessary factorisation $K(\alpha)=K_{+}(\alpha) K_{-}(\alpha)$ is discussed in  Appendix~\ref{appendix_a}. 

For the decomposition \eqref{f_K} it is necessary  to remove the pole $\alpha=-p_n^{(1)}$ from $F_{-}(\alpha)$. In such way  one obtains
\begin{equation}
 F_{-}(\alpha)=f_1(\alpha) \left ( \frac{1}{K_{-}(\alpha)}-\frac{1}{K_{-}(-p_n^{(1)}) }\right ), \qquad F_{+}(\alpha)=\frac{f_1(\alpha)}{K_{-}(-p_n^{(1)})}\, .
\end{equation}
Using  \eqref{WH_solutions} and  \eqref{phi_minus} leads to the following expressions
\begin{equation}
A(\alpha)=\frac{C_1(\alpha)}{\sin\big (q(\alpha) h\big )},\qquad B(\alpha)=\frac{C_1(\alpha)}{\sin\big (q(\alpha) (b-h)\big )}
\end{equation}
where
\begin{equation}
C_1(\alpha)=-\frac{i \sin (\pi nh/b )K_{-}(\alpha)}{\big (\alpha+p_n^{(1)}\big )K_{-}(-p_n^{(1)})\sqrt{bp_n^{(1)}}}\, .
\end{equation}

\subsection{Second solution}
For these solution the incident wave $\Psi_{\mathrm{inc}}=\phi_{n}^{(2)-}(x,y)$ (see \eqref{the_second}). Therefore from \eqref{zero_h} it follows that 
\begin{equation}
\Phi_{+}(\alpha,h)=0
\label{phi_plus_zero}
\end{equation}
and, consequently,  
\begin{equation}
X_{-}(\alpha)\equiv \Phi_{-}(\alpha,h)=A(\alpha)\sin\big (q(\alpha)h\big )=B(\alpha)\sin\big (q(\alpha)(b-h)\big ).
\label{A_B_second}
\end{equation}
The incident field  $\phi_{n}^{(2)-}(x,y)$ is continuous at $y=h$, $-\infty<x<0$ but its $y$ derivative has a jump at this line. Therefore,  $\Phi_{-}(\alpha,y)$ is also continuous at this line but its $y$-derivative should compensate the discontinuity of  $y$ derivative of the incident field (cf. \eqref{y_prime_h}). It means that 
\begin{equation}
\Phi_{-}^{\prime}(\alpha,h+0)-\Phi_{-}^{\prime} (\alpha,h-0)=-\frac{i \pi (-1)^n n }{(b-h)\big (\alpha-p_n^{(2)}\big )\sqrt{(b-h)p_n^{(2)}}}\, . 
\end{equation}
Comparing the $y$-derivatives from the both sides of $y=h$ with \eqref{A_B} gives the following equation
\begin{eqnarray}
&&X_{+}(\alpha)\equiv b\left ( \Phi_{+}^{\prime}(\alpha,h+0)-\Phi_{+}^{\prime}(\alpha,h-0)\right )=\\
&=&\frac{i \pi (-1)^n n\, b }{(b-h)\big (\alpha-p_n^{(2)}\big )\sqrt{(b-h)p_n^{(2)}}}-b q(\alpha) 
\left ( B(\alpha) \cos\big (q(\alpha)(b-h)\big )+A(\alpha) \cos\big (q(\alpha)h\big )\right )\,.\nonumber
\end{eqnarray}
Expressing $A(\alpha)$ and $B(\alpha)$ from \eqref{A_B_second} leads to the Wiener-Hopf equation \eqref{WH_equation} with $K(\alpha)$  given by  \eqref{K_alpha} and 
\begin{equation}
f_2(\alpha)=\frac{i \pi (-1)^n n b K(\alpha)}{(b-h)\big (\alpha-p_n^{(2)}\big )\sqrt{(b-h)p_n^{(2)}}}\, .
\end{equation}
Solving the resulting Wiener-Hopf equation  equation and using  \eqref{A_B_second} one gets
\begin{equation}
A(\alpha)=\frac{C_2(\alpha)}{\sin\big (q(\alpha) h\big )},\qquad B(\alpha)=\frac{C_2(\alpha)}{\sin\big (q(\alpha) (b-h)\big )} 
\end{equation}
where
\begin{equation}
C_2(\alpha)=\frac{i \pi(-1)^n nb \, K_{+}(p_n^{(2)})K_{-}(\alpha)}{(b-h)\big (\alpha-p_n^{(2)}\big )\sqrt{(b-h)p_n^{(2)}}}\, .
\end{equation}

\subsection{Third solution}
In this case the incident field $\Psi_{\mathrm{inc}}=\phi_{n}^{(3)-}(x,y)$ and Eqs.~\eqref{phi_plus_zero} and \eqref{A_B_second} remain valid. 

From discontinuity of the incident field one gets
\begin{equation}
\Phi_{-}^{\prime}(\alpha,h+0)-\Phi_{-}^{\prime} (\alpha,h-0)=-\frac{i(-1)^n \pi n }{h\big (\alpha-p_n^{(3)}\big )\sqrt{hp_n^{(3)}}}.
\end{equation}
As above one obtains the same Wiener-Hopf equation but with 
\begin{equation}
f_3(\alpha)=\frac{i(-1)^n \pi n b K(\alpha)}{h\big (\alpha-p_n^{(3)}\big )\sqrt{hp_n^{(3)}}}\, .
\end{equation}
In this case one finds 
\begin{equation}
A(\alpha)=\frac{C_3(\alpha)}{\sin \big (q(\alpha) h\big )},\qquad B(\alpha)=\frac{C_3(\alpha)}{\sin\big (q(\alpha) (b-h)\big )}
\end{equation}
with
\begin{equation}
C_3(\alpha)=\frac{i(-1)^n \pi n b\, K_{+}(p_n^{(3)})K_{-}(\alpha)}{h\big (\alpha-p_n^{(3)}\big )}\, . 
\end{equation}
\subsection{$S$-matrix for irrational  $h/b$} 
Finding  the reflection field by the inverse Fourier transform \eqref{inverse_fourier} and using the relation $K_{-}(-\alpha)=K_{+}(\alpha)$ after simple but tedious calculations  one gets the explicit form of the $S$-matrix
\begin{equation}
S_{n,m}^{\alpha\to \beta}=\frac{L_n^{(\alpha)} L_m^{(\beta)}}{x_n^{(\alpha)}+x_m^{(\beta)}}
\label{S_matrix}
\end{equation}
where
\begin{equation}
L_n^{(1)}=\frac{\sin\big (\pi n h/b\big )}{ K_{+}(p_n^{(1)})\sqrt{bp_n^{(1)}}},\quad 
L_n^{(2)}= -\frac{(-1)^n\pi n  b \, K_{+}(p_n^{(2)})}{(b-h)\sqrt{(b-h)p_n^{(2)}}},\quad 
L_n^{(3)}= -\frac{(-1)^n  \pi n  b\,  K_{+}(p_n^{(3)})}{h \sqrt{h p_n^{(3)}}}
\label{L_n}
\end{equation}
and
\begin{equation}
x_n^{(1)}=bp_n^{(1)},\qquad x_n^{(2)}=-bp_n^{(2)},\qquad x_n^{(3)}=-bp_n^{(3)}.
\end{equation}
Function $K_{+}(\alpha)$ is given by (convergent) infinite product \eqref{K_plus} and values of momenta $p_m^{(\beta)}$ with $\beta=1,2,3$  are indicated in \eqref{elementary_solutions}. 

Notice that the $S$-matrix is symmetric as it should be from the reciprocity principle. 

The above expressions  are  valid  when the ratio $h/b$ is an irrational number  which means that (i)  $\sin\big (\pi h n/b\big )\neq 0$ for all $n$  and (ii) all $p_m^{(\alpha)}$ are different.  For rational $h/b$ certain terms in the above formulas will have formally uncertainties of $0/0$ type. Though they  can be resolved by taking a corresponding limit it is more convenient to reconsider the resonance case separately. 

\subsection{$S$-matrix for rational $h/b$}\label{resonant_S}

Let, for simplicity,   $h/b=1/q$ with integer $q$. In this case the following  3 moments are equal 
\begin{equation}
p_n^{(3)}=p_{(q-1)n}^{(2)}=p_{qn}^{(1)}
\end{equation}
and it is plain that  waves $\psi^{(1)\pm}_{qn}(x,y)$
are two exact solutions of the problem considered as they are automatically zero on the full line passing through the barrier.  As 
\begin{equation}
\frac{1}{\sqrt{bp_{qn}^{(1)}}}\sin\left  (\frac{\pi q n}{b}y\right ) =\left \{ \begin{array}{cc} \frac{1}{\sqrt{h p_{n}^{(3)}}}\sin\Big (\frac{\pi  n}{h}y\Big)\sqrt{\frac{h}{b}}, &0<y<h\\
 \frac{1}{\sqrt{(b-h) p_{(q-1)n}^{(2) }}}\sin\Big (\frac{\pi (q-1)n}{b-h}(b-y)\Big)(-1)^{nq+1}  \sqrt{\frac{b-h}{b}},&h<y<b
 \end{array}\right .
 \label{exact_solution} 
\end{equation}
it follows without calculations  that 
\begin{equation}
S_{n, qn}^{3\to 1}=\frac{1}{\sqrt{q}}\, ,\qquad  S_{(q-1)n, qn}^{2\to 1}=-(-1)^{qn} \sqrt{1-\frac{1}{q}}\, . 
\label{resonance}
\end{equation}
These and the symmetric counterparts are the only elements which are formally singular at rational $h/b$. Other matrix elements may be obtained directly from the preceding Section. 

For barrier billiard as in figure~\ref{barrier}(a) with rational $h/b$ the above plane waves are the exact eigenfunctions with simple eigenvalues. Therefore it is convenient to remove them when non-trivial spectrum  is considered.    Function $\psi^{(1)+}_{qn}(x,y)$ can be removed by simply removing the term with $m=qn$ for all possible $n$. To remove  $\psi^{(1)-}_{qn}(x,y)$ it is necessary to form a special combinations of the incident waves $\psi^{(2)-}_{(q-1) n}(x,y)$ and $\psi^{(3)-}_{n}(x,y)$ such that there is no scattering into $\psi^{(1)-}_{qn}(x,y)$
From \eqref{resonance} it follows that the correct combination which cancels this exact solution at negative $x$ is 
\begin{equation}
\phi^{(2+3)}_{n(q-1)}=\frac{1}{\sqrt{q}}\phi^{(2)}_{n(q-1)}+(-1)^{nq} \sqrt{\frac{q-1}{q}} \phi_n^{(3)}  
\label{correct_combination}
\end{equation}
Notice that it is a symbolic notation. It just means that in the second region one has a function  $\frac{1}{\sqrt{q}}\phi^{(2)}_{n(q-1)}$ and in the third region in such case one should consider a function  $(-1)^{nq} \sqrt{\frac{q-1}{q}} \phi_n^{(3)}$.
For convenience we label this function by indices from the second region. 

Using the above formulas for the $S$-matrix one finds that $h/b=1/q$ the $S$-matrix with excluded exact solutions has the same  form as in \eqref{S_matrix} 
\begin{equation}
S_{n,m}^{\alpha\to \beta}=\frac{L_n^{(\alpha)} L_m^{(\beta)}}{x_n^{(\alpha)}+x_m^{(\beta)}}
\end{equation}
where indices $\alpha,\beta=1,2,2+3$. $L_n^{(1)}$ and $L_n^{(2)}$ are as in \eqref{L_n} but with the above restrictions on admissible values of $n$
\begin{eqnarray}
L_n^{(1)}&=&\frac{\sin(\pi n h/b)}{ K_{+}^{\mathrm{res}}(p_n^{(1)})\sqrt{bp_n^{(1)}}},\qquad n\neq 0 \mod q ,\\
L_n^{(2)}&=& -\frac{(-1)^n\pi n  b K_{+}^{\mathrm{res}}(p_n^{(2)})}{(b-h)\sqrt{(b-h)p_n^{(2)}}}, \qquad n\neq 0 \mod q-1
\end{eqnarray}
 but 
\begin{equation}
L_{n}^{(2+3)}=-\frac{\pi n (-1)^n b^{3/2} K_{+}^{\mathrm{res}}(p_n^{(2)})}{ (b-h)^{3/2} \sqrt{h p_n^{(2)}}},\qquad n=0 \mod q-1.
\label{resonant_L_3}
\end{equation}
Function $K_{+}^{\mathrm{res}}(\alpha)$ is the value of $K_{+}(\alpha)$ in \eqref{K_plus} when $h/b=1/q$ and its value  is given by \eqref{resonant_K}. 
 
The expressions in the previous and this Sections are exact. Together with \eqref{B_matrix},  \eqref{exact_phases}, and \eqref{determinant}  they represent an exact surface-of-section quantisation of the considered barrier billiards. They can also serve for numerical calculations  in these models.  But the purpose  of the paper is to extract from the exact solution a random matrix whose spectral statistics in the semiclassical limit $k\to\infty$ will be the same as spectral statistics of barrier billiard. This is discussed in the next Section.


\section{Main random matrix}\label{unitary_random_matrix}

\subsection{Restriction to propagating modes}

Formally the considered $S$-matrix is infinite. It is related with the existence of two types of waves, the propagating modes for which the momentum is real and evanescent modes with imaginary momentum.  The number of propagating modes in each channel is finite
\begin{equation}
N_1=\left [ \frac{kb}{\pi} \right ],\qquad N_2=\left [ \frac{k(b-h)}{\pi} \right ],\qquad N_3=\left [ \frac{kh}{\pi} \right ]
\label{numbers_modes}
\end{equation}
but the number of evanescent modes is infinite. 

 In  the  semiclassical limit $k\to\infty$ evanescent modes  decrease  quickly from the barrier tip and  they are negligible for high exited states  provided that the barrier tip is not at  distances of the order of wavelength  from the billiard boundaries. When evanescent modes with imaginary momenta are ignored the $S$-matrix as well as the $B$-matrix become final $N\times N$  unitary matrices with $N=N_1+N_2+N_3$. 
 
From general principles of quantum mechanics the $S$-matrix for propagating modes has to be unitary.
Till now all quantities have  2 indices: a subscript indicated quantum number and a superscript descrying to what channel belongs this quantum number. It is convenient to organise indices of propagating modes into one super index $\mu$ from $1$ to $N$ with the convention that indices from $1$ to $N_1$ belong to the first channel, indices from $N_1+1$ to $N_2$ to the second channel and indices from $N_1+N_2+1$ till $N=N_1+N_2+N_3$ indicate the third channel. It is also useful  
 to combine  3 vectors  of propagating modes $bp^{(j)}_m$, $j=1,2,3$ into one vector $x_{\mu}$ of dimension $N$ such that 
\begin{equation}
\vec{x}=b\Big( \underbrace{ p_1^{(1)},\ldots,  p_{N_1}^{(1)}}_{\text{$N_1$}}, \underbrace{-p_1^{(2)},\ldots, - p_{N_2}^{(2)}}_{\text{$N_2$}},\underbrace{-p_1^{(3)},\ldots,  -p_{N_3}^{(3)}}_{\text{$N_3$}}\Big ).
\label{vector_x}
\end{equation}
Notice that the second and the third parts have minus sign. 

In such notations the scattering $S$-matrix takes a simple form
\begin{equation}
S_{\mu,\nu}=\frac{L_{\mu} L_{\nu}}{x_{\mu}+x_{\nu}},\qquad \mu,\nu=1,\ldots, N
\label{general_S_matrix}
\end{equation}
which is exactly the form of matrix obtained in \cite{bogomolny} but for different specification of $\vec{x}$. 

As indicated in this paper, the unitary condition $SS^{\dag}=1$  (and the Cauchy determinant formula) implies that 
\begin{equation}
|L_{\mu}|^2=2x_{\mu} \prod_{\nu\neq \mu} \frac{x_{\mu}+x_{\nu}}{x_{\mu}-x_{\nu}}
\label{L_square}
\end{equation}
In Appendix~\ref{appendix_a} it is checked that these relations are consequences of the $S$-matrix expressions discussed in the preceding Sections. 

As quantities $|L_{\mu}|^2$ have to be positive there exists a general condition on set $x_{\mu}$, $\mu=1,\ldots,N$ \cite{bogomolny}.  If the moduli of $x_{\mu}$ are ordered
\begin{equation}
|x_1|>|x_2|>\ldots >|x_{N}|
\label{ordered_x}
\end{equation}
then the positivity of $|L_m|^2$ implies that 
\begin{equation}
x_{\mu}=(-1)^{\mu+1}|x_{\mu}|,\qquad \mu=1,\ldots,N 
\label{intertwining} 
\end{equation}
Elementary arguments demonstrate that vector \eqref{vector_x} obeys these intertwining conditions.  

In the  case when $h/b=1/q$ and exact modes are removed the $S$-matrix has the same form as in \eqref{general_S_matrix} and \eqref{L_square}  but coordinates $x_{\mu}$ with $\mu=1,2,\ldots, N$ with  dimension
\begin{equation}
 N= N_1+N_2-N_3 
 \label{N_1_q}
 \end{equation}
has to be arranged into the following vector
\begin{equation}
\vec{x} =b\Big( \underbrace{ p_1^{(1)},\ldots,p_k^{(1)},\ldots,  p_{N_1}^{(1)}}_{\text{$k\neq 0 \mod q$}}, \underbrace{-p_1^{(2)},\ldots, - p_{N_2}^{(2)}}_{\text{$N_2$}} \Big ). 
\label{x_rational}
\end{equation}
Here it is implicitly assumed that terms $p_m^{(2)}$ with $m\neq $ mod $q-1$ belong to the second channel and terms $p_m^{(2)}$ with $m=0 $ mod $q-1$ form $2+3$ channel discussed above  with $L_{\mu}^{(2+3)}$ given by \eqref{resonant_L_3}. Notice that for the symmetric barrier billiard  with $q=2$ this result coincides with the one obtained in \cite{bogomolny}. 

In the general case when the ratio $h/b=m/q$ with co-prime integers $m$ and $q$ ($m<q$) there are 3 degenerated momenta 
\begin{equation}
p_{qt}^{(1)}=p_{(q-m)t}^{(2)}=p_{mt}^{(3)}
\end{equation} 
with integer $t$. 

The same arguments as above show that when the simple exact solutions are removed  coordinate vector $\vec{x}$ can be chosen in the following form
\begin{equation}
\vec{x}=b\Big( \underbrace{ p_1^{(1)},\ldots,p_k^{(1)},\ldots,  p_{N_1}^{(1)}}_{\text{$k\neq 0 \mod q$}}, \underbrace{-p_1^{(2)},\ldots, - p_{N_2}^{(2)}}_{\text{$N_2$}},\underbrace{-p_1^{(3)},\ldots,-p_k^{(3)},\ldots,   -p_{N_3}^{(3)}}_{\text{$k\neq 0 \mod m $}}\Big )\, .
\label{x_m_q}
\end{equation}
The dimension of this vector is 
\begin{equation}
N=N_1+N_2+N_3-2N_0,\qquad 
 N_0=\left [\frac{kb}{\pi q} \right]\, .
 \label{N_m_q}
 \end{equation}
   

\subsection{Random phase approximation}

When in the semiclassical limit $k\to\infty$ evanescent modes  are ignored the $S$-matrix and, consequently,  the $B$-matrix  \eqref{B_matrix}  become finite dimensional unitary matrices.  The $B$-matrix plays the role of the transfer operator and its  importance  lies in the fact that  spectral statistics of this matrix is (up to a rescaling)  the same as spectral statistics of the barrier billiard (see \cite{semiclassics, DS} and a short discussion in \cite{bogomolny}). 

The $B$-matrix \eqref{B_matrix} differs from the scattering $S$-matrix by phase factors  $e^{i\phi_m}$ where $\phi_m$ given by \eqref{exact_phases} depend on momenta and horizontal sizes of the barrier billiard.   On can argue \cite{bogomolny} that in semiclassical limit $k\to \infty$ these phases  for propagating modes can be considered as independent random variables uniformly distributed on the unit circle.  Though the proof of this random phase approximation  is unknown to the author, physically it is quite natural from the following considerations:   
\begin{itemize}
\item $\phi_m$ are non-linear functions of $m$,
\item $\phi_m \to \infty$ in semiclassical limit $k\to \infty$,
\item $\phi_m$ with different $m$ are non-commensurable (when $h/b$ is an irrational number or when $h/b$ is rational and exact solutions are removed).
\end{itemize} 
The random phase approximation, in particular, states that local spectral correlation functions of deterministic phases  \eqref{exact_phases} mod $2\pi$ after unfolding should be the same as for the Poisson distribution of independent uniformly distributed random variables, namely
\begin{equation}
P_n(s)=\frac{s^{n-1}}{(n-1)!} e^{-s}
\label{poisson}
\end{equation}
where $P_n(s)$ with $n=0,1,\ldots$  is the probability density that two variables are separated  by distance $s$ and there exit exactly $n$ other variables inside this interval. 
 
\begin{figure}
\begin{center}
\includegraphics[width=.7\linewidth]{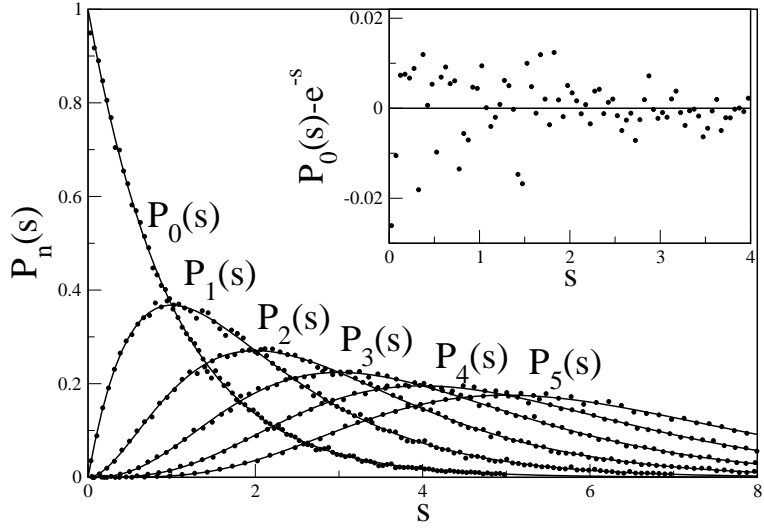}
\end{center}
\caption{Nearest neighbour distributions for deterministic phases \eqref{exact_phases} with $k=50\,000.5$, $b=\pi$, $h=b/\sqrt{5}$, $a=2$, $d_1=1$, $d_2=1.5$.  Solid lines are the Poisson predictions for these quantities \eqref{poisson}.  Insert: the difference between numerically calculated $P_0(s)$ and $\exp(-x)$. }
\label{general_phases}
\end{figure}

For illustration, in figure~\ref{general_phases} the six  nearest neighbour distributions for deterministic phases \eqref{exact_phases} with indicated parameters are plotted. In total $N=99\,999$ numbers (cf., \eqref{numbers_modes}) were taken into account. It is clearly seen that calculated correlation functions agree well with the Poisson predictions which gives a certain credit to the random phase approximation but, of course, cannot prove it.  The absence of analytical confirmations of this approximation implies  that the results below should be considered valid for 'typical' barrier billiards or for 'almost all values' of billiard parameters without giving a precise definition.  The situation is, in a sense, similar to the physical statement that  eigenvalues of 'generic'  quantum integrable systems are well  described by the Poisson statistics \cite{berry_tabor}. Even for a rectangular billiard where all eigenvalues are known explicitly  one can establish rigorously  only that (for a certain ratio of the sides) the  two-point correlation function agrees  with the Poisson value \cite{marklof}. To prove that the nearest-neighbour distributions in this case are close to the Poisson expressions  \eqref{poisson} (which is well confirmed by numerics) seems to be beyond the existing methods. 

Taken the random phase approximation as granted the unitary random matrix associated with the considered barrier billiards takes the form 
\begin{equation}
B_{\mu,\nu}=e^{i\Phi_{\mu} } \frac{L_{\mu} L_{\nu}}{x_{\mu}+x_{\nu}} , \qquad \mu,\nu=1,\ldots,N\ .
\label{random_B_matrix}
\end{equation}
Quantities $L_{\mu}$ are, in general,  complex numbers, $L_{\mu}=|L_{\mu}|e^{i\xi_{\nu}}$  whose moduli $|L_{\mu}|$ are  related with $x_\mu$  by simple expressions \eqref{L_square} being merely  a consequence of the unitarity. On the other hand, phases of $L_{\mu}$  are nontrivial and given by an infinite product as in  Appendix~\ref{appendix_a}.  By a conjugation phases of $L_{\nu}$ can be shifted to the first factor $L_{\mu}$ and now the total phase factor takes the form $\Phi_{\mu}^{\prime} =\Phi_{\mu}+2\xi_{\mu}$.  As  $\Phi_{\mu} $ are assumed to be  independent random variables uniformly distributed at the unit circle the same will be   valid for  $\Phi_{\mu}^{\prime}$.  These arguments establish that in the definition \eqref{random_B_matrix} coefficients $L_{\mu}$ can be considered as  real  numbers 
\begin{equation}
L_{\mu}=\sqrt{2x_{\mu} \prod_{\nu\neq \mu} \frac{x_{\mu}+x_{\nu}}{x_{\mu}-x_{\nu}}}
\end{equation} 
provided that vector $\vec{x}$ obeys the intertwining conditions \eqref{ordered_x} and \eqref{intertwining}. 

The random unitary matrix $B$ \eqref{random_B_matrix} depends on two groups of parameters. The first consists on $N$ independent random phases $\Phi_m$ uniformly distributed between $0$ and $2\pi$. The second includes $N$ coordinates $x_{\mu}$ obeying \eqref{ordered_x} and \eqref{intertwining}. To describe the barrier billiard indicated in figure~\ref{barrier}  with irrational ratio $h/b$ vector $x_{\mu}$ has to chosen as indicated in \eqref{vector_x} and when this ratio is rational and the exact modes are  removed it has the form \eqref{x_rational} or \eqref{x_m_q}.  

Though the $B$-matrix has been extracted from the exact quantisation of barrier billiards it remains meaningful  for arbitrary  coordinate  sequence $x_{\mu}$ (obeying the above intertwining conditions).  

The $B$-matrix \eqref{random_B_matrix} (as well as many other matrices with intermediate statistics) belongs to the so-called class of low complexity matrices \cite{complexity} with a displacement operator of low rank \cite{displacement} (cf.,  \cite{bogomolny}). The exact correlation functions for such matrices, in general,  are unknown.  It has been conjectured  in \cite{toeplitz} that nearest neighbour distributions  for such matrices are well described by the normalised gamma-distributions 
\begin{equation}
P_n(s)=a_n s^{\gamma_n}e^{-b_n s}, \qquad \int_0^{\infty}P_n(s)ds=1,\qquad \int_0^{\infty}s P_n(s)ds=n+1
\end{equation}
depended only on one parameter $\gamma_n$ which can be calculated as follows
\begin{equation}
\gamma_n=q_n-1
\end{equation}
with $q_n$ equal the minimal number of  parameters (co-dimension) needed to get exactly $n+2$ degenerated eigenvalues.  

For Hermitian matrices  the value of co-dimension is given by a theorem by von Neumann and  Wigner and its generalisations \cite{neumann_wigner, keller}.  For low complexity matrices and, in particular, for the $B$-matrix discussed above no exact results about $q_n$ are available (see \cite{toeplitz} for 'physical' determination of this quantity).   It was argued in \cite{bogomolny} that irrespective of the choice of coordinates $x_{\mu}$, the co-dimension $q_n$  for the unitary $B$-matrix equals $2n+2$ and, therefore,  
 $\gamma_n=2n+1$. It means that  the nearest neighbour distributions for random $B$-matrices should be well described by the following Wigner-type surmise 
\begin{equation}
P_n(s)=\frac{2^{2n+2} }{(2n+1)!} s^{2n+1} e^{-2s}
\label{semi_poisson} 
\end{equation}
which corresponds to the so-called semi-Poisson distribution \cite{plasma_model}. 

\subsection{Numerical calculations}

\begin{figure}
\begin{center}
\includegraphics[width=.7\linewidth]{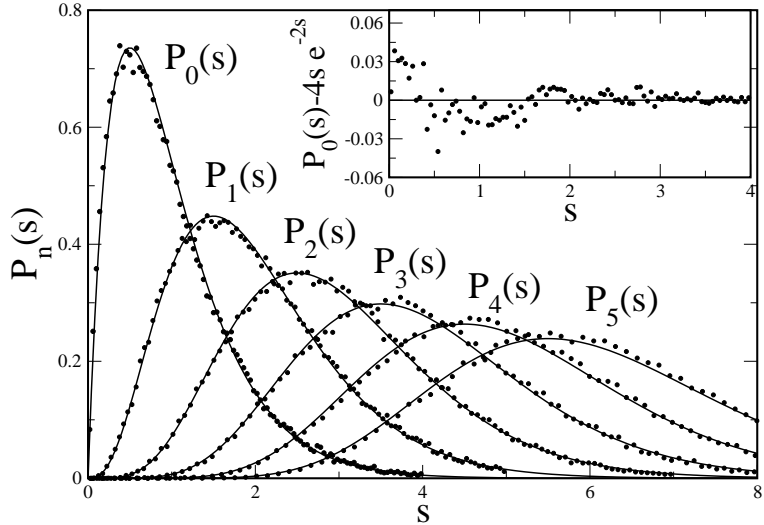}
\end{center}
\caption{Nearest-neighbour distributions for random $N\times N$ $B$-matrices with irrational ratio   $h=b/\sqrt{5}$ (black circles). In calculations $N=999$ corresponded to $k=500.5$, $b=\pi$, and $100$ realisations of random phases. Solid lines are the semi-Poisson formulas \eqref{semi_poisson}. Insert: the difference between numerically calculated 
$P_0(s)$ and the semi-Poisson prediction: $4se^{-2s}$.}
\label{general_barrier_sqrt(5)}
\end{figure}

\begin{figure}
\begin{center}
\includegraphics[width=.7\linewidth]{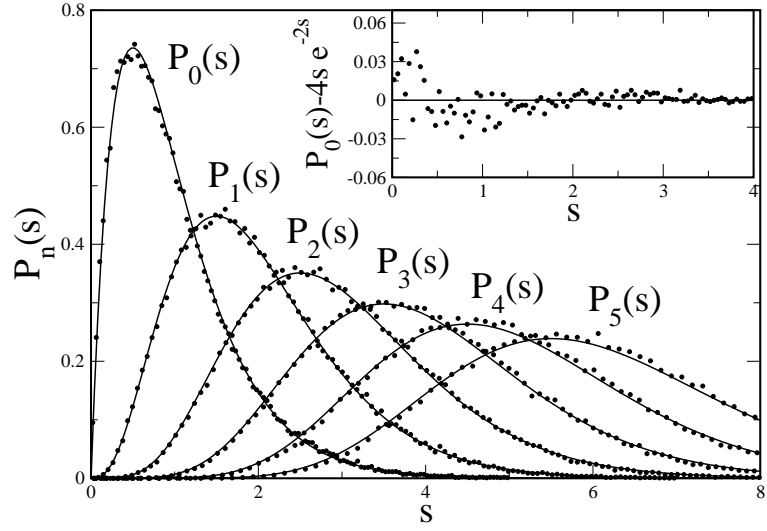}
\end{center}
\caption{The same as in figure~\ref{general_barrier_sqrt(5)} but for ratio $h/b=1/3$,  $k=800.5$, and  $N=1067$. }
\label{general_barrier_3}
\end{figure}

\begin{figure}
\begin{center}
\includegraphics[width=.7\linewidth]{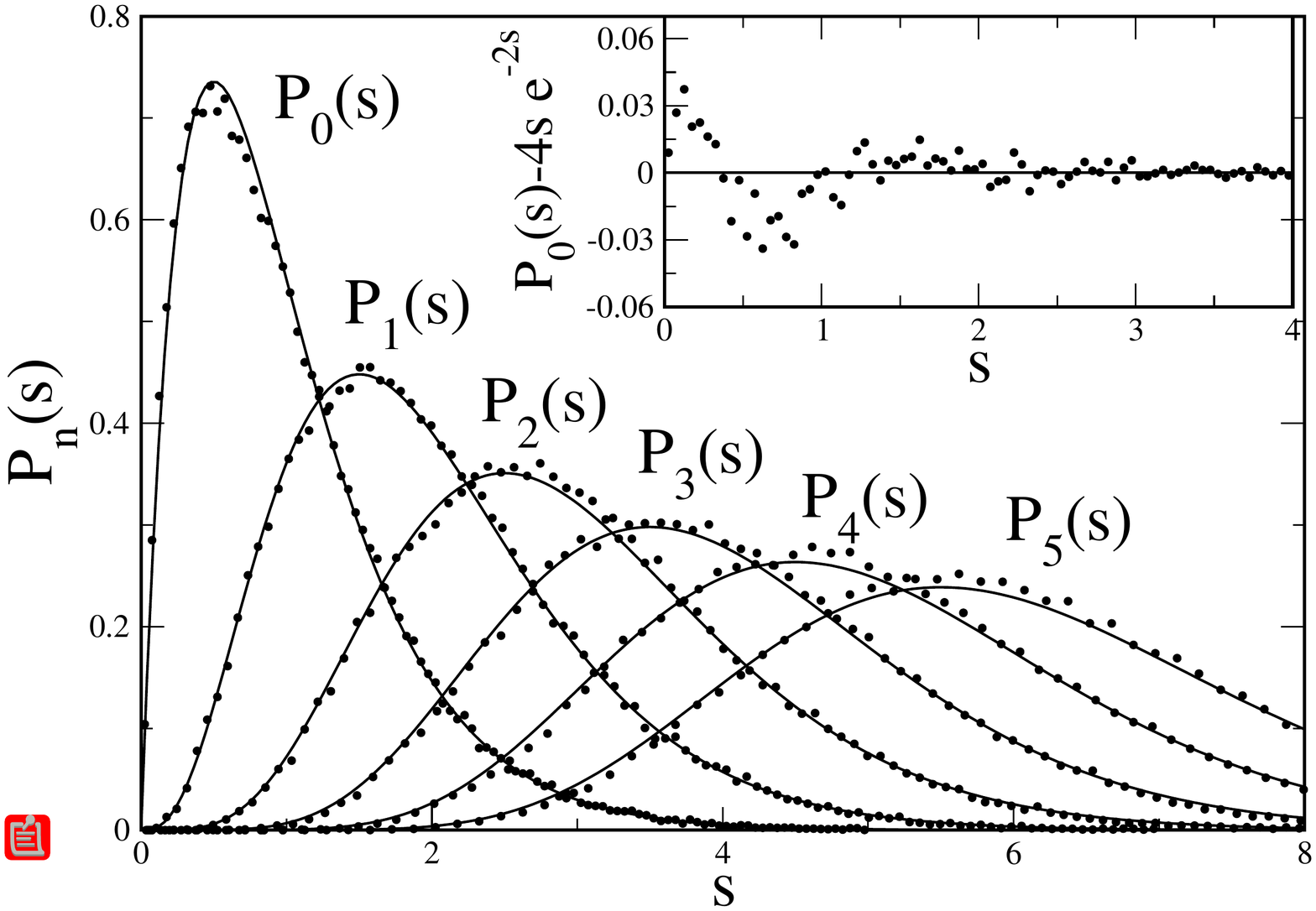}
\end{center}
\caption{The same as in figure~\ref{general_barrier_sqrt(5)} but for ratio $h/b=2/5$, $k=650.5$, and  $N=1040$. }
\label{general_barrier_2_5}
\end{figure}

To check these predictions numerical calculations of local spectral statistics for the above $B$-matrices with  irrational and rational ratios $h/b$ were performed. Typical results are presented in figures~\ref{general_barrier_sqrt(5)},  \ref{general_barrier_3}, and \ref{general_barrier_2_5}. The first figure shows 6 lowest nearest-neighbour distributions for the random $B$-matrix with $h/b=1/\sqrt{5}$, i.e., with $x_{\mu}$ given by  \eqref{vector_x} with $b=\pi$, $h=\pi /\sqrt{5}$,  and $k=500.5$ which  according to \eqref{numbers_modes} leads to matrix dimension $N=999$. In the second figure the ratio $n/b$ was chosen equal to $1/3$ and the momentum $k=800.5$ which gives matrix dimension $N=1067$ (cf., \eqref{N_1_q}). In the third figure $h/b=2/5$ and $k=650.5$ which corresponds to  $N=1040$ (see \eqref{N_m_q}). In all figures the results are averaged over  $100$ realisations of random phases. It is clearly seen that simple semi-Poisson formulas \eqref{semi_poisson} described quite well the numerical results.

\section{Summary}\label{summary}

Barrier billiards are simple  examples of pseudo-integrable systems with not-trivial and interesting classical and quantum properties. The main result of the paper is the construction of unitary random matrices associated with  general barrier billiards.  The principal importance of these matrices is that their spectral statistics is the same (up to a rescaling)  as spectral statistics of two-dimensional barrier billiards. 

To extract such random matrices from the exact quantum description of these billiards the following steps were done.    
\begin{itemize}
\item Billiard boundaries perpendicular to the barrier were removed.
\item The resulting  problem  corresponds to the scattering in an infinite slab with a half-plane inside.
\item The exact $S$-matrix for this problem was calculated by the Wiener-Hopf method. 
\item In such a way one gets exact functions obeying  both  the Helmholtz equation and the correct boundary conditions on boundaries parallel to the barrier. 
\item Proper billiard eigenfunctions were written as linear combinations  of these scattering waves with coefficients being obtained from boundary conditions on boundaries perpendicular to the barrier. 
\item  The exact quantisation condition of barrier billiards appears in  the form $\det(1+B(k))=0$ where the $B$-matrix differs from the $S$-matrix by phase factors depended on billiard dimensions parallel to the barrier. 
\item In the semiclassical limit two main simplifications occur:\\
(i)   $S$ and $B$ matrices become finite dimensional unitary matrices after ignoring exponentially small evanescent modes.\\
(ii) The deterministic phases of propagating modes  are substituted by independent random phases uniformly distributed on the unit circle. 
\end{itemize}
 The obtained  $N\times N$ random unitary $B$-matrix depend on  2 groups of parameters: $N$ random phases and $N$ coordinates obeyed certain intertwining conditions. To describe barrier billiards coordinates have to be specially specified (see \eqref{vector_x}, \eqref{x_rational}, \eqref{x_m_q}) but the the $B$-matrix remains meaningful for an arbitrary choice of coordinates.  
 
 The spectral statistics of these $B$-matrices belongs to intermediate-type statistics which differs from both the Poisson statistics typical for integrable systems and the usual random matrix statistics characteristic for chaotic systems.  Heuristic calculation of the Wigner-type surmise for these matrices reveals that their local spectral statistics is well described by the semi-Poisson distribution in a good agreement with numerical calculations. 
Further  investigation of statistical properties of the $B$-matrices will  be done elsewhere. 

Applying these results to the barrier billiards leads to the following conclusions 
\begin{itemize}
\item Local spectral correlation functions for rectangular billiards with a barrier inside as in figure~\ref{barrier}(a) remain the same for (almost) all positions and heights of the barrier. 
\item These correlation functions are well described by the semi-Poisson distribution.
\item The same should be also true for more general  barrier billiards indicated in figure~\ref{barrier}(b) with  irrational ratio $h/b$.  The case of these billiards with rational ratio $h/b$ requires a special investigation which is beyond the score of this paper.
\end{itemize}
Numerical calculations of spectral correlation functions  were done so far only for symmetric barrier billiard \cite{wiersig,gorin_wiersig} and they are in agreement with  the above statements. For more general barrier billiards discussed in the paper no numerics is known to the author. Analytical calculations of the spectral compressibility for general barrier billiards as in figure~\ref{barrier}(a) with irrational ratio $d/a$ and $h/b=m/q$ with $m,q$ being co-prime integers were done in \cite{olivier_general}. The result of this paper reads
\begin{equation}
K(0)=\frac{1}{2}+\frac{1}{q}
\end{equation} 
but  in the calculations exact eigenstates discussed in Section~\ref{resonant_S} were not removed. When these modes are  put off the result is $K(0)=1/2$ \cite{private} which coincides with the semi-Poisson prediction and corroborates the results of the paper.


\appendix
\section{Factorisation of $K(\alpha)$}\label{appendix_a}

The purpose of this Appendix is to calculate the  factorisation of the kernel  $K(\alpha)$  into the product $K_{+}(\alpha) K_{-}(\alpha)$ needed in the application of the Wiener-Hopf method in Sections~\ref{non_resonant_S} and \ref{resonant_S}. 

For irrational ratio $h/b$ function $K(\alpha)$ \eqref{K_alpha}  has the following form
\begin{equation}
K(\alpha)=\frac{\sin\big (q(\alpha)h\big )\sin\big (q(\alpha)(b-h)\big )}{b q(\alpha)\sin\big (q(\alpha)b\big )}
\end{equation}
Each sinus function in this expression gives rise to the following series of zeros   with integer $n\geq 1$ ($q(\alpha)=0$ is not a singular point) 
\begin{equation}
q(\alpha^{(\beta)})=\frac{\pi n}{h_{\beta}}, \qquad \alpha^{(\beta)}=\pm \sqrt{k^2-\frac{\pi^2 n^2}{h_{\beta}^2}}\equiv \pm p_n^{(\beta)}, \qquad \beta=1,2,3
\end{equation}
where $h_{\beta}$  is the width of the corresponding channel  given by \eqref{h_beta}.

Function $K_{+}(\alpha)$  has to be free of singularities in the upper half-plane of complex $\alpha$. Therefore it should include  only negative zeros and poles. The needed factors can easily be calculated from the well-known  formula
\begin{equation}
\sin x=x\prod_{n=1}^{\infty}\left (1-\frac{x^2}{\pi^2 n^2}\right ).
\label{prod_sinus}
\end{equation}
It is plain that 
\begin{equation}
K_{+}(\alpha)=\frac{\sqrt{h(b-h)}}{b}\prod_{m=1}^{\infty} \frac{\Big ( \sqrt{1-k^2b_m^{(2)2}}-i\alpha b_m^{(2)}\Big )\Big (\sqrt{1-k^2b_m^{(3)2}}-i\alpha b_m^{(3)}\Big )}{\Big (\sqrt{1-k^2b_m^{(1)2}}-i\alpha b_m^{(1)} \Big )}
\label{K_plus}
\end{equation}
where $b_m^{(\beta)}=h_{\beta}/(\pi m)$.

For the case $h/b=1/q$ considered in Section~\ref{resonant_S}  
 $b_n^{(3)}=b_{nq}^{(1)}$ and the infinite product in  \eqref{K_plus} contains only two factors
\begin{equation}
K_{+}^{\mathrm{res}}(\alpha)=\frac{\sqrt{h(b-h)}}{b} \frac{\prod_{m=1 }^{\infty} \Big  ( \sqrt{1-k^2b_m^{(2)2}}-i\alpha b_m^{(2)}\Big  )}{\prod_{\substack{m=1\\ m\neq 0 \mod q} }^{\infty}\Big  (\sqrt{1-k^2b_m^{(1)2}}-i\alpha b_m^{(1)} \Big )}\, .
\label{resonant_K} 
\end{equation}

From general considerations the scattering $S$-matrix has to be unitary for propagating modes which implies the validity of  \eqref{L_square}. It is instructive to check this fact directly from the above formulas. The main step is the calculation   $|K_{+}(p_n^{(\beta)})|^2$ for $\beta=1,2,3$.  Even for propagating modes this function includes both  the product of  finite number of propagating modes with real momenta and the product of  infinite number of evanescent modes which have  pure imaginary momenta.  Let us calculate them  separately for each factor in \eqref{K_plus} without explicitly bothering about the convergence.  

Define
\begin{equation}
w^{(\beta)}(\alpha)=\prod_{m=1}^{\infty} \Big ( \sqrt{1-k^2b_m^{(\beta)2}}-i\alpha b_m^{(\beta)}\Big ),\qquad \beta=1,2,3\ .
\end{equation}
One has for real $\alpha$ (with $N_{\beta}$ from \eqref{numbers_modes})
\begin{equation}
w^{(\beta)}(\alpha)=\prod_{m=1}^{N_{\beta}} \Big (-i \sqrt{k^2b_m^{(\beta)2}-1}-i\alpha b_n^{(\beta)}\Big )\prod_{m=N_{\beta}+1}^{\infty} \Big ( \sqrt{1-k^2b_m^{(\beta)2}}-i\alpha b_m^{(\beta)}\Big )\, .
\end{equation}
Therefore
\begin{equation}
\Big |w^{(\beta)}(\alpha)\Big |^2=\prod_{m=1}^{N_{\beta}} b_m^{(\beta) 2}\Big ( \alpha+p_m^{(\beta)} \Big)^2\prod_{m=N_{\beta}+1}^{\infty}\Big (1-(k^2-\alpha^2)b_m^{(\beta) 2} \Big ) \, . 
\end{equation}
Assume that $\alpha= p_n^{(l)}$ with $l\neq \beta$. Then the second product is not zero for all $m$ and one gets
\begin{equation}
\Big |w^{(\beta)}(p_n^{(l)})\Big |^2=\prod_{m=1}^{N_{\beta}} b_m^{(\beta) 2}\frac{\Big ( p_n^{(l)}+p_m^{(\beta)} \Big)^2}{\Big (1- (h_{\beta} n)^2/(h_{l} m)^2 \Big )}
\prod_{m=1}^{\infty}\Big (1- \frac{h_{\beta}^2 n^2}{h_{l}^2 m^2} \Big )\,  . 
\label{intermediate}
\end{equation}
Using the fact that $1- (h_{\beta}n)^2/(h_{l} m)^2=b_m^{(\beta)2}\big ( p_n^{(l)2}-p_m^{(\beta)2}\big )$
and taking into account \eqref{prod_sinus} one gets that
\begin{equation}
\Big |w^{(\beta)}(p_n^{(l)})\Big |^2=\frac{h_l\sin (\pi h_{\beta} n/h_l)}{\pi h_{\beta} n} 
 \prod_{m=1}^{N_{\beta}} \frac{ p_n^{(l)}+p_m^{(\beta)} }{p_n^{(l)}-p_m^{(\beta)}}, \qquad  l\neq \beta\,  .
\end{equation}
For $l=\beta$ this transformation  does not work as the second product  in \eqref{intermediate} is zero for $m=n$. This formal difficulty  can easily be avoid by the following manipulation (taking into account that $n$ is from a propagating mode and $m$ is the second product is from an evanescent mode) 
\begin{equation}
\prod_{m=N_{\beta}+1}^{\infty} \Big (1- \frac{ n^2}{m^2} \Big )=\frac{\prod_{m=1, \ m\neq n}^{\infty} \Big (1- \frac{n^2}{m^2} \Big )}{\prod_{m=1, \ m\neq n}^{N_{\beta} } \Big (1- \frac{ n^2}{ m^2} \Big )} \, .
 \end{equation}
 The calculation of the numerator is performed as follows
 \begin{equation}
 \prod_{m=1, \ m\neq n}^{\infty} \Big (1- \frac{n^2}{m^2} \Big )=\lim_{x\to n} \frac{1}{1-x^2/n^2} \prod_{m=1}^{\infty} \Big (1- \frac{x^2}{m^2} \Big )=\lim_{x\to n} \frac{\sin \pi x}{\pi x(1-x^2/n^2)}=-\frac{1}{2} (-1)^n\, .
 \end{equation}
 Therefore
 \begin{equation}
 \Big |w^{(\beta)}(p_n^{(\beta)})\Big |^2=
 \frac{2(-1)^{n+1} p_n^{(\beta)2} h_{\beta}^2}{\pi^2  n^2} \prod_{m=1,\ m\neq n}^{N_{\beta}} \frac{ p_n^{(\beta)}+p_m^{(\beta)} }{p_n^{(\beta)}-p_m^{(\beta)}} \, .
 \end{equation}
 Finally one finds that
 \begin{eqnarray}
\Big  |K_{+}(p_n^{(1)})\Big |^2&=&\frac{\sin^2(\pi h n/b)}{2 b^2 p_n^{(1)2}}\prod_{m=1,\ m\neq n}^{N_1}\frac{p_n^{(1)}-p_m^{(1)}}{p_n^{(1)}+p_m^{(1)}} \prod_{m=1}^{N_2}\frac{p_n^{(1)}+p_m^{(2)}}{p_n^{(1)}-p_m^{(2)}}\prod_{m=1}^{N_3}\frac{p_n^{(1)}+p_m^{(3)}}{p_n^{(1)}-p_m^{(3)}}\, , 
 \label{mod_K_1}\\
\Big  |K_{+}(p_n^{(2)})\Big |^2&=&-\frac{2  (b-h)^3 p_n^{(2)2 }}{b \pi^2 n^2}\prod_{m=1}^{N_1}\frac{p_n^{(2)}-p_m^{(1)}}{p_n^{(2)}+p_m^{(1)}} \prod_{m=1,\ m\neq n}^{N_2}\frac{p_n^{(2)}+p_m^{(2)}}{p_n^{(2)}-p_m^{(2)}}\prod_{m=1}^{N_3}\frac{p_n^{(2)}+p_m^{(3)}}{p_n^{(2)}-p_m^{(3)}}\, ,  \label{mod_K_2}\\
\Big  |K_{+}(p_n^{(3)})\Big |^2&=&-\frac{2  h^3 p_n^{(3)2 }}{b \pi^2 n^2}\prod_{m=1}^{N_1}\frac{p_n^{(3)}-p_m^{(1)}}{p_n^{(3)}+p_m^{(1)}} \prod_{m=1}^{N_2}\frac{p_n^{(3)}+p_m^{(2)}}{p_n^{(3)}-p_m^{(2)}}\prod_{m=1,\ m\neq n}^{N_3}\frac{p_n^{(3)}+p_m^{(3)}}{p_n^{(3)}-p_m^{(3)}} \, .
 \end{eqnarray}
Taking into account \eqref{L_n} it is plain  that the relations \eqref{L_square} implying the unitarity  are fulfilled. 
  

\end{document}